\documentclass[twocolumn,english,aps,groupedaddress,prl]{revtex4}

\usepackage[colorlinks=true,allcolors=blue]{hyperref}
\usepackage{array}
\usepackage[latin9]{inputenc}
\usepackage{amsmath}
\usepackage{amssymb}
\usepackage{graphicx}
\usepackage{babel}
\usepackage{mathrsfs}
\usepackage{amsfonts}
\usepackage{epstopdf}
\usepackage{tabularx}
\usepackage{multirow}
\usepackage{color}
\usepackage{natbib}
\usepackage{bm}
\usepackage{amsthm}
\usepackage{siunitx}
\usepackage{braket}
\usepackage{hhline}

\newcommand{\ba}{\mathbf{a}}

\newcommand{\be}{\mathbf{e}}

\newcommand{\br}{\mathbf{r}}

\usepackage{xcolor}

\begin{document}

\title{Universal Spin-Position Coupled Rydberg Interactions}
\author{Chengshu Li}
\email{chengshu@mail.tsinghua.edu.cn}
\author{Hui Zhai}
\email{hzhai@tsinghua.edu.cn}
\affiliation{Institute for Advanced Study, Tsinghua University, Beijing 100084, China}
\affiliation{Beijing Key Laboratory of Cold Atom Quantum Computation, Tsinghua University, 100084, Beijing, China}

\date{\today}

\begin{abstract}

Strong interactions between $s$-orbital Rydberg atoms underpin atom-array-based quantum computation and quantum simulation. Typically, the spin dependence of these interactions is negligible and plays no practical role. In this Letter, we uncover that a strong electron-spin-dependent Rydberg interaction can emerge when the principal quantum numbers of the two $s$-orbital atoms differ by a sweet-spot value. This interaction originates from the fine-structure splitting of nearby $p$ orbitals, which endows it with distinct symmetry properties such that the electron spins are coupled to the relative position of the two atoms, featuring a spatially defined anisotropy. We further demonstrate that this interaction admits a universal form, independent of the principal quantum numbers, and highlight its fundamental distinction from conventional magnetic dipolar interactions, establishing it as a new type of native magnetic interaction in nature. Our findings introduce a new element to the Rydberg quantum simulation toolbox. As a concrete application, we propose a native realization of the Kitaev--Heisenberg model, which hosts an unusual stripe phase as a quantum many-body manifestation of spin--space locking.

\end{abstract}

\maketitle

Rydberg atom arrays have become firmly established as indispensable platforms for quantum simulation and quantum computation~\cite{Morgado2021,Saffman2010}. This success relies on precise laser control of atomic states, geometrical reconfigurability via optical tweezers, and, most critically, the strong interactions between Rydberg atoms. These interactions underpin the realization of two-qubit CZ gates~\cite{Levine2019,Fu2022,Jandura2022,Bluvstein2022,Evered2023} and enable the discovery of novel quantum many-body phases, including quantum many-body scars~\cite{Bernien2017,Turner2018a,Turner2018b,Zhai,Chandran2023} and quantum spin liquids~\cite{Verresen2021,Semeghini2021,Giudici2022,Tarabunga2022,Cheng2023}.

Current mainstream quantum computation and simulation tasks primarily utilize $s$-orbital Rydberg states. In these systems, the interactions depend essentially on the Rydberg-state occupation and are largely insensitive to the specific electronic spin state that the atom occupies. This spin insensitivity originates from the nature of the Rydberg interaction, namely, the van der Waals interaction, which arises as a second-order process of the electric dipolar coupling. The relevant energy scales are governed by the dipolar interaction strength and the energy gap between the $s$-orbital and nearby $p$-orbitals. As long as the electronic spin--orbit coupling, characterized by the fine-structure splitting, remains much smaller than these dominant scales, its effects can be safely neglected, leaving the electron spin degree of freedom effectively decoupled from the Rydberg interactions.

Nevertheless, a careful \textit{ab initio} calculation of the Rydberg interaction consistently reveals some degree of electronic spin dependence~\cite{Gallagher1994,Shi2014,Weber2017,Mogerle2026}. However, this aspect has never been theoretically quantified. Outstanding questions include whether there exist regimes where such spin dependence becomes pronounced, and whether it admits an analytical and universal form. In this Letter, we address these questions and uncover remarkably simple and general answers. Specifically, there exists a sweet-spot difference in the principal quantum numbers for which the fine-structure splitting becomes comparable to the intermediate-state energy gap, thereby giving rise to a strong electron-spin-exchange coupling between the two atoms. Furthermore, symmetry considerations alone yield a universal form for this spin-dependent interaction. Owing to its origin in spin-orbit coupling, this interaction exhibits a distinct anisotropy that locks the spin and spatial degrees of freedom, naturally lending itself to quantum simulation tasks such as the realization of the Kitaev--Heisenberg model.

A number of previous works have focused on spin-exchange interactions in Rydberg atom arrays~\cite{Browaeys2016,deLeseleuc2019,Chen2023,Emperauger2025,Yue2026,Qiao2025}. In all these studies, however, the two spin components represent pseudo-spins associated with different orbital degrees of freedom. This scenario differs fundamentally from the one we discuss here in several key aspects. First, the pseudo-spin exchange does not rely on fine-structure splitting. Second, it lacks spatial locking and therefore does not exhibit the spin--space-locked anisotropy. Third, its symmetry properties are entirely different. As we will detail below, the electron-spin exchange interaction possesses a rotation symmetry that simultaneously rotates both spin and spatial coordinates, whereas the pseudo-spin exchange is inherently of the XXZ type~\cite{Browaeys2016,deLeseleuc2019,Chen2023,Emperauger2025,Yue2026}, where the $z$-direction is singled out by the orbital degree of freedom, making it an intrinsically special axis.

\begin{figure}[t]
    \centering
    \includegraphics[width=0.85\linewidth]{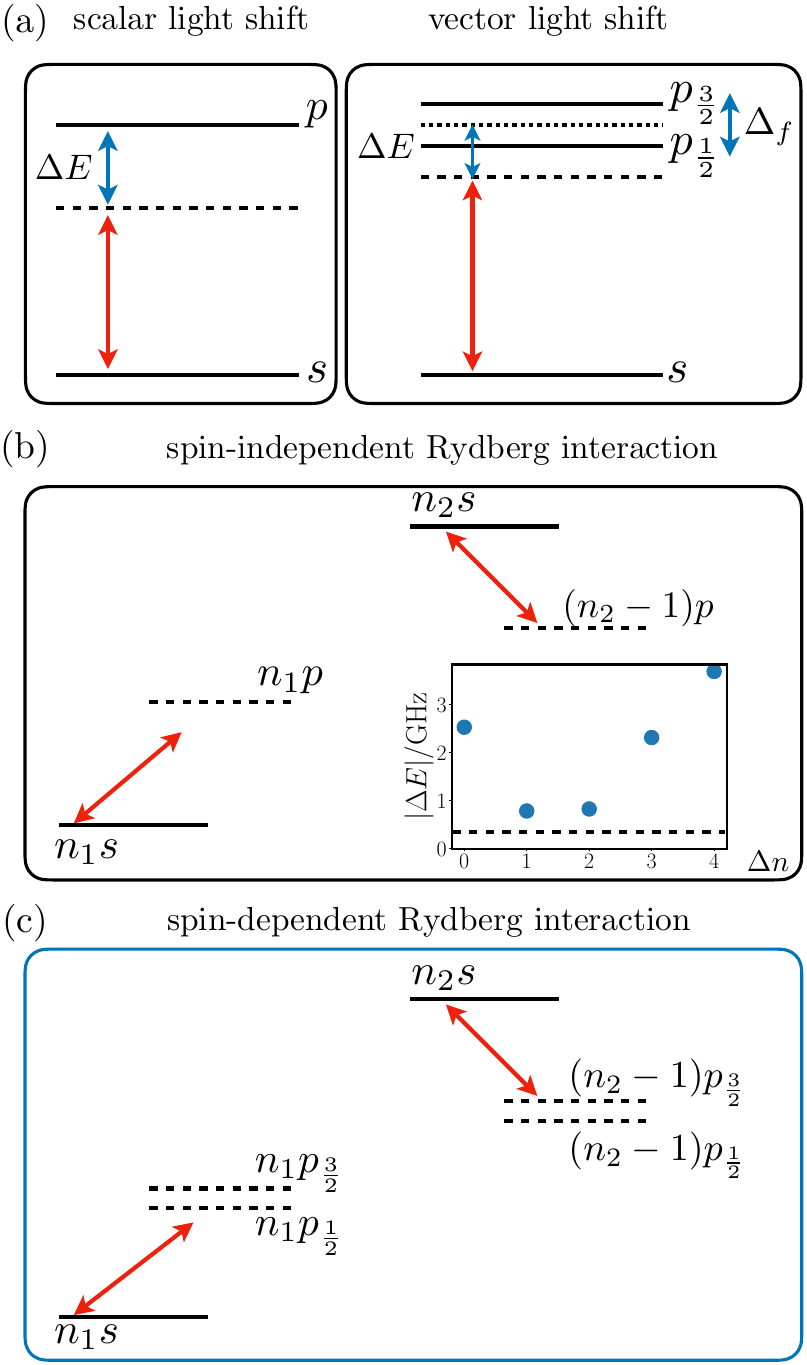}
    \caption{From light shift to Rydberg interactions. (a) When a laser drives the atom, the detuning $\Delta E$ and the spin--orbit splitting $\Delta_f$ define two competing energy scales. Depending on whether $\Delta E \gg \Delta_f$ or $\Delta E \sim \Delta_f$, the induced light shift becomes either spin-independent (scalar) or spin-dependent (vector). (b) In conventional Rydberg interactions, the electric dipole--dipole coupling acts as a perturbation, and the relevant energy defect $\Delta E = E_{n_1,p}+E_{n_2-1,p}-E_{n_1,s}-E_{n_2,s}$ is typically much larger than $\Delta_f$. The inset shows $\Delta E$ as a function of $\Delta n = n_1 - n_2$. (c) When $\Delta E$ becomes comparable to $\Delta_f$, the fine-structure splitting of the intermediate states must be explicitly taken into account, resulting in spin-dependent Rydberg interaction.}
    \label{fig:energy}
\end{figure}

\emph{When does electron spin become important in Rydberg interaction?} 

To develop physical intuition, we first recall a well-known phenomenon in atomic physics, namely, the light shift. This shift arises from a second-order perturbative process mediated by the electric dipole coupling between the atom and the light field, with a detuning energy $\Delta E$, as illustrated in Fig.~\ref{fig:energy}(a). When $\Delta E$ greatly exceeds the fine-structure splitting $\Delta_f$ of the $p$ orbitals, the spin--orbit coupling can be safely neglected; consequently, the electron spin is decoupled, yielding a purely scalar light shift. In the opposite regime, where $\Delta E$ becomes comparable to $\Delta_f$, the finite fine-structure splitting must be explicitly included in the calculation. This leads to a spin-dependent vector light shift, which has been extensively exploited for quantum simulating spin--orbit coupling in ultracold atomic gases~\cite{Zhai2015,Zhai2021}.

We now generalize this single-particle picture to the two-body Rydberg interaction. In this case, the role of the external light field is replaced by the electric dipole--dipole interaction between atoms. Despite this difference, the underlying coupling remains purely electric in nature. Consequently, any pronounced dependence on the electron spin can only enter through the intermediate states.

Let us consider two Rydberg atoms initially occupying the $s$ orbitals with principal quantum numbers $n_1$ and $n_2$, respectively. The most significant perturbation channel, illustrated in Fig.~\ref{fig:energy}(b), proceeds through the $n_1,p$ and $(n_2-1), p$ orbitals, as these give the largest dipole matrix elements and the smallest energy defects. The intermediate state energy difference
\begin{equation}
\Delta E=E_{n_1,p}+E_{n_2-1,p}-E_{n_1,s}-E_{n_2,s},
\end{equation}
can be estimated using quantum defect theory, which yields $E_{n,l,j}=-E_0/(n-\delta_l)^2$ with $E_0=13.6\text{ eV}$. As a rough estimate, we neglect the $n$ and $j$ dependence of the quantum defect $\delta_l$ and set $\delta_s=3.1$ and $\delta_p=2.6$. For example, in a typical experimental realization with $n \simeq 60$, $|\Delta E|$ reaches its minimum at $n_2-n_1=2$~\cite{deltan1}, as shown in the inset of Fig.~\ref{fig:energy}(b). One can further verify that this $\Delta E$ is already comparable to typical values of the fine-structure splitting. Therefore, analogous to the vector light-shift scenario, for these chosen principal quantum numbers, the fine-structure splitting must be explicitly included in the second-order perturbative calculation, as shown in Fig.~\ref{fig:energy}(c). Consequently, a significant electron-spin dependence emerges in the Rydberg interaction. Indeed, \textit{ab initio} calculations show that for $n_1-n_2=2$, the spin-dependent interaction strength is about two orders of magnitude larger than its spin-independent counterpart, and is comparable to the spin-independent Rydberg interaction strength for $n_1-n_2=0$.

\emph{How does the Rydberg interaction depend on the electronic spins?} 

\begin{figure}
    \centering
    \includegraphics[width=\linewidth]{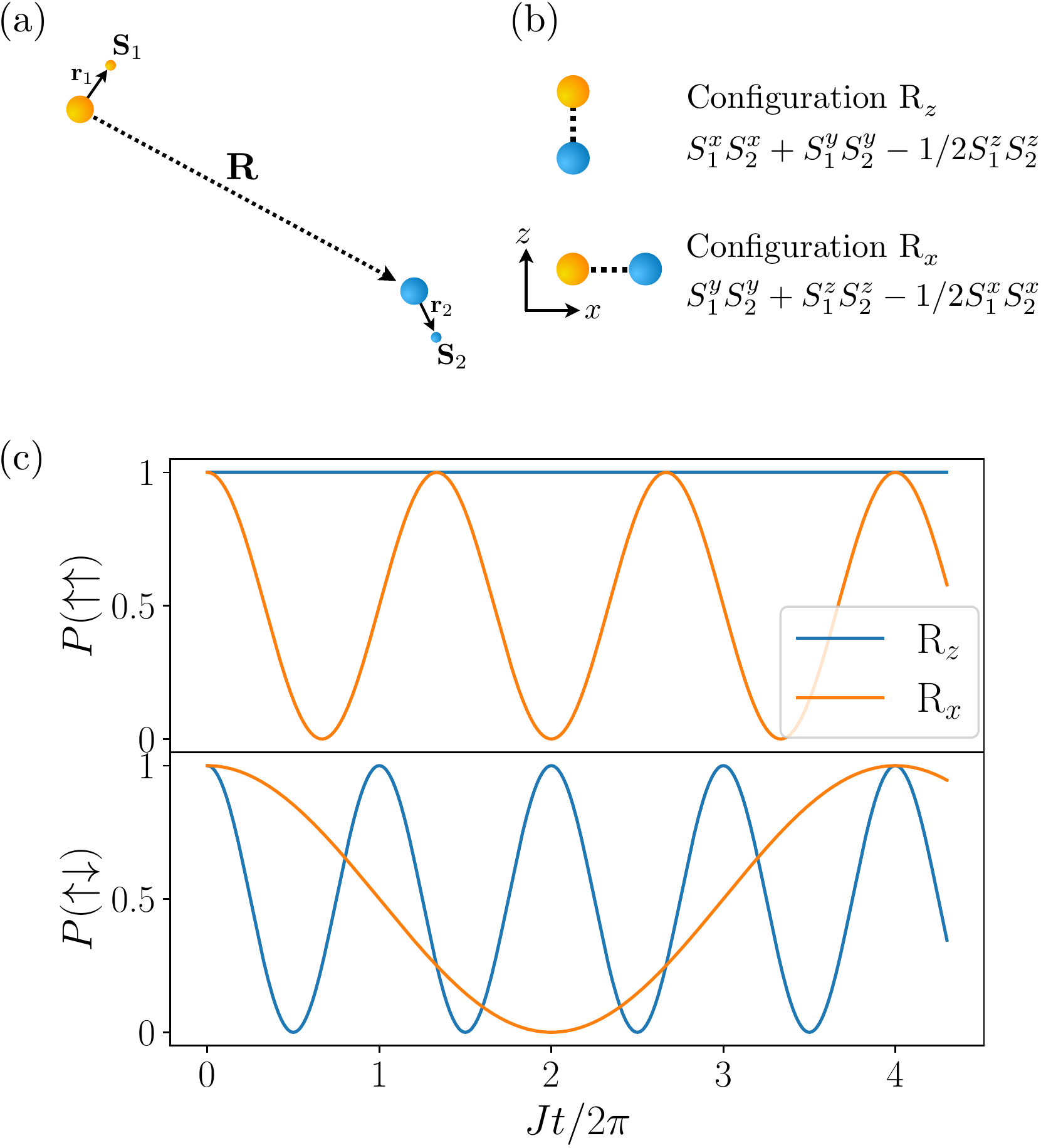}
   \caption{Symmetry analysis and experimental consequences. (a) The spin-dependent Rydberg interaction involves two physical mechanisms: the electric dipole--dipole interaction between atoms and the electron spin--orbit coupling within each atom. (b) Two configurations with distinct spin dynamics. The spin quantization axis is along $\hat{z}$, and the interatomic displacement is either parallel to $\hat{z}$ (R$_z$ configuration) or to $\hat{x}$ (R$_x$ configuration). (c) Spin dynamics for two initial states $\ket{\uparrow\uparrow}$ and $\ket{\uparrow\downarrow}$ in the two configurations (R$_z$ and R$_x$).}
    \label{fig:symmetry}
\end{figure}

That said, the spin-dependent part of the Rydberg interaction arises from the cooperative interplay between the electric dipole--dipole interaction and the electron spin--orbit coupling. Before proceeding to detailed calculations, we can determine the universal form of this interaction from the following symmetry considerations, as illustrated in Fig.~\ref{fig:symmetry}(a):

\begin{itemize}
  \item The spin--orbit coupling locks each atom's electron spin, denoted by $\mathbf{S}_i$ ($i=1,2$), to the relative orbital angular momentum $\mathbf{L}_i$ of its electron around the nucleus. Consequently, the system does not possess independent rotation symmetry of each electron spin; instead, only simultaneous rotations of the electron spin and the associated spatial motion around the nucleus are allowed.
  
  \item The electron dipole moment is $\mathbf{d}_i = e\,\mathbf{r}_i$, where $\mathbf{r}_i$ is the relative displacement between the electron and its nucleus. Therefore, any rotation of the electron's orbital motion directly rotates $\mathbf{d}_i$.
  
  \item The dipole--dipole interaction reads
  \begin{equation}
  H_\mathrm{dd}=\frac{1}{4\pi\epsilon_0 R^3}
  \Big(\mathbf{d}_1\cdot\mathbf{d}_2 - 3(\mathbf{d}_1\cdot\hat{\mathbf{R}})(\mathbf{d}_2\cdot\hat{\mathbf{R}})\Big),
  \end{equation}
  where $\hat{\mathbf{R}}=\mathbf{R}/R$, $\mathbf{R}$ is the relative displacement between the two atoms, and $R=|\mathbf{R}|$. This interaction implies that the rotation symmetry must simultaneously act on both $\mathbf{r}_i$ ($i=1,2$) and $\mathbf{R}$.
  
  \item The total interaction is invariant under time-reversal symmetry, i.e., it remains unchanged under $\hat{\mathbf{S}}_1 \rightarrow -\hat{\mathbf{S}}_1$ and  $\hat{\mathbf{S}}_2 \rightarrow -\hat{\mathbf{S}}_2$.
\end{itemize}

Combining the first three considerations, we find that the resulting spin-dependent Rydberg interaction preserves only a single rotation symmetry, namely, the simultaneous rotation of $\mathbf{S}_1$, $\mathbf{S}_2$, and $\hat{\mathbf{R}}$. Therefore, the allowed bilinear terms are $\mathbf{S}_1\cdot\mathbf{S}_2$, $\mathbf{S}_1\cdot\hat{\mathbf{R}}$, and $\mathbf{S}_2\cdot\hat{\mathbf{R}}$. The last symmetry consideration forbids the latter two terms individually, but allows their product $(\mathbf{S}_1\cdot\hat{\mathbf{R}})(\mathbf{S}_2\cdot\hat{\mathbf{R}})$. Moreover, since the Rydberg interaction is a second-order process, it depends on $\mathbf{S}_i$ at most quadratically. Hence, we arrive at the universal form
\begin{equation}
H = J\Big(\mathbf{S}_1\cdot\mathbf{S}_2 + \alpha\,(\mathbf{S}_1\cdot\hat{\mathbf{R}})(\mathbf{S}_2\cdot\hat{\mathbf{R}})\Big) + \cdots, \label{spin-dependent-int}
\end{equation}
where $J$ denotes the overall strength of the spin-dependent part, and $\cdots$ represents the spin-independent contribution. The coefficient $\alpha$ must be obtained by including all four physical processes shown in Fig.~\ref{fig:energy}(c) and calculating the energy defects and dipole matrix elements of the intermediate states. It turns out that the Wigner--Eckart theorem dictates that $\alpha$ is purely geometric, independent of the principal quantum numbers $n_1$ and $n_2$, and further calculation shows that $\alpha$ can be straightforwardly evaluated as $-3/2$.

We note that the interaction in Eq.~\eqref{spin-dependent-int} differs fundamentally from the magnetic dipolar interaction in several key aspects. First, its strength is comparable to that of the Rydberg blockade when the interatomic distance is on the order of a few micrometers. This is because the interaction is electric in origin, making it orders of magnitude stronger than typical magnetic dipolar couplings. Second, the coefficient $\alpha$ in the magnetic dipolar case is $+3$, in contrast to $\alpha=-3/2$ found here. Finally, the magnetic dipolar interaction decays as $1/R^3$, while the present interaction follows a $1/R^6$ spatial decay, characteristic of van der Waals-type interactions.

The spin-space locking encoded in Eq.~\eqref{spin-dependent-int} can be directly probed in experiments. Suppose we prepare the initial state as an eigenstate of the spin quantization axis $\hat{z}$, and consider two geometries, R$_z$ and R$_x$, in which the two atoms are aligned parallel to the $z$ and $x$ axes, respectively, as shown in Fig.~\ref{fig:symmetry}(b). For demonstration, we work in the zero-Zeeman-energy limit. We prepare the atoms in either $\ket{\uparrow\uparrow}$ or $\ket{\uparrow\downarrow}$, and monitor the time evolution in the $z$-basis. For the R$_z$ configuration, the interaction takes the form $S^x_1S^x_2+S^y_1S^y_2-\frac{1}{2}S^z_1S^z_2$, which possesses a $U(1)$ rotation symmetry about the $\hat{z}$ axis. This symmetry ensures that $\ket{\uparrow\uparrow}$ remains stationary, while $\ket{\uparrow\downarrow}$ undergoes exchange oscillations with frequency $J$. For the R$_x$ configuration, the interaction becomes $-\frac{1}{2}S^x_1S^x_2+S^y_1S^y_2+S^z_1S^z_2$, under which both initial states exhibit nontrivial time evolution, with characteristic frequencies $3J/4$ and $J/4$, respectively, as shown in Fig.~\ref{fig:symmetry}(c). In practice, the spin quantization axis is typically defined by an external magnetic field, and the finite Zeeman energy will modify these frequencies. Nevertheless, the qualitative difference in dynamics between the two geometries remains the definitive experimental signature of this spin-dependent Rydberg interaction.

\emph{What is the spin--position locked Rydberg interaction good for?}

\begin{figure}
    \centering
    \includegraphics[width=\linewidth]{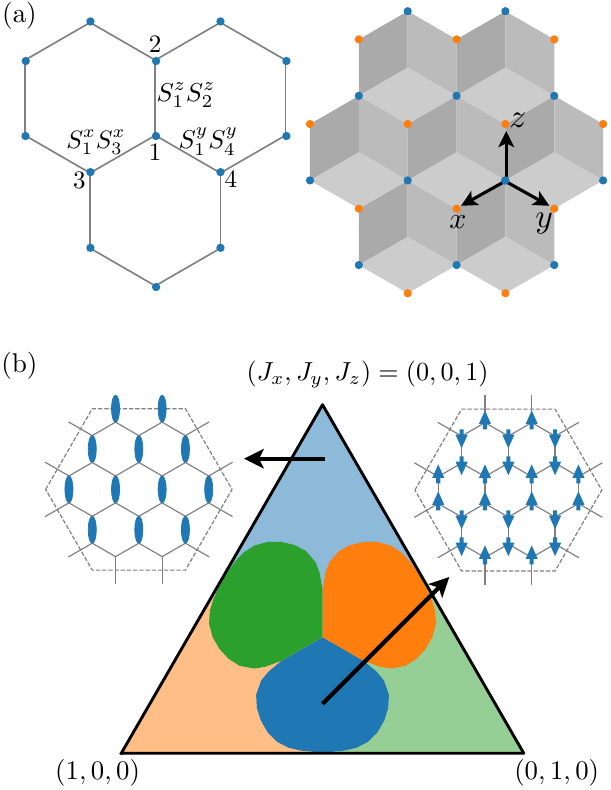}
    \caption{Realizing the Kitaev--Heisenberg model. (a) Left: Schematic of the Kitaev model, where the spin interactions are of the form $S_i^\gamma S_j^\gamma$ with $\gamma=x,y,z$ on the three distinct bond types. Right: Proposed quantum simulation of the Kitaev--Heisenberg model with $\alpha=-3/2$. The atoms are arranged into two two-dimensional layers, with nearest-neighbor bonds aligned along the $x$, $y$, and $z$ axes. The shaded surfaces are included to visualize the underlying three-dimensional cubic stacking, which provides an intuitive picture of the atomic positions. (b) Phase diagram as a function of the relative interaction strengths among the three bonds. The dark orange, green, and blue regions denote stripe phases with spin ordering along the $x$, $y$, and $z$ directions, respectively, while the light orange, green, and blue regions correspond to singlet-product phases with singlet formation on the $x$, $y$, and $z$ bonds. The phase boundaries are obtained from exact diagonalization on a 24-site cluster.}
    \label{fig:2d}
\end{figure}

The spin--space-locked interaction that naturally emerges from our scheme provides a new resource for quantum simulation. As a concrete example, we show that this interaction naturally realizes the Kitaev--Heisenberg model. The Kitaev honeycomb lattice model is a celebrated paradigm of spin--space-locked anisotropy: on each site, the three bonds are associated with spin interactions of the form $S^x_i S^x_j$, $S^y_i S^y_j$, and $S^z_i S^z_j$, respectively (see Fig.~\ref{fig:2d}(a)). This exactly solvable model serves as a prototypical platform hosting both gapped and gapless quantum spin liquid phases~\cite{Kitaev2006}. Since its proposal, the search for Kitaev physics in quantum materials, such as in certain iridate compounds~\cite{Chaloupka2010,Takagi2019,Matsuda2025}, has remained a central scientific goal. In realistic materials, the Heisenberg exchange terms are generally present alongside the Kitaev coupling. Consequently, the resulting Kitaev--Heisenberg model has attracted considerable attention in quantum magnetism over the past two decades~\cite{Osorio2014,Sela2014,Oitmaa2015,Gohlke2017,Catuneanu2019,Dong2020,Zou2025}.

We propose that the Kitaev--Heisenberg Hamiltonian can be naturally realized by arranging atoms in two layers of two-dimensional planes, as shown in Fig.~\ref{fig:2d}(a). Atoms with the two distinct principal quantum numbers are placed on the blue ($a$) and orange ($b$) sites, respectively, with their positions given by
\begin{equation}
\begin{split}
&\br_a(n_1,n_2)=n_1\ba_x+n_2\ba_y,\\
&\br_b(n_1,n_2)=\ba_0+n_1\ba_x+n_2\ba_y,    
\end{split}
\end{equation}
where $\ba_0=a_z(0,0,1)$, $n_{1,2}\in\mathbb{Z}$, the basis vectors are $\ba_x=(a_x,-a_y,0)$ and $\ba_y=(0,a_y,-a_z)$, and $a_{x,y,z}$ denote the lattice constants along the $x$, $y$, and $z$ directions. Defining also $\ba_z=(-a_x,0,a_z)$, we note that the system has translational symmetry along $\ba_{x,y,z}$. It is evident that all blue sites lie in one two-dimensional plane and all orange sites in another, a configuration that can be readily implemented with optical tweezer arrays. Moreover, the nearest-neighbor links, e.g., between $\br_a(n_1,n_2)$ and each of $\br_b(n_1,n_2)$, $\br_b(n_1,n_2+1)$, and $\br_b(n_1+1,n_2+1)$, are displacement vectors parallel to the $x$, $y$, and $z$ axes, with lengths $a_x$, $a_y$, and $a_z$, respectively. An intuitive way to visualize this geometry is to introduce an auxiliary cubic lattice with sites at $(n_1a_x, n_2a_y, n_3a_z)$; the two atom types reside on two $(a_x^{-1}, a_y^{-1}, a_z^{-1})$-planes separated by $\sqrt{a_x^2+a_y^2+a_z^2}/3$. The shaded surfaces of the cubic lattice in Fig.~\ref{fig:2d}(a) illustrate this three-dimensional structure. Consequently, the nearest-neighbor spin-dependent interactions are given by Eq.~\eqref{spin-dependent-int}, with $\mathbf{R}$ taking the values $a_x\be_x$, $a_y\be_y$, and $a_z\be_z$ along the three corresponding nearest-neighbor bonds.

Since the Rydberg interaction decays rapidly with increasing interatomic distance, we restrict ourselves to nearest-neighbor couplings. With this setup, the system naturally realizes the Kitaev--Heisenberg model
\begin{equation}
H=\sum_{\langle ij\rangle} J_\gamma \left(\mathbf{S}_i\cdot\mathbf{S}_j + \alpha\, S_i^\gamma S_j^\gamma\right),
\end{equation}
where $\alpha = -3/2$ is fixed by the microscopic derivation, and $J_\gamma \propto a_\gamma^{-6} > 0$ ($\gamma = x,y,z$) for the experimentally relevant regime. Crucially, the relative strengths of $J_x$, $J_y$, and $J_z$ can be independently tuned by adjusting the lattice constants $a_\gamma$. Previous theoretical studies have shown that in the isotropic limit $J_x=J_y=J_z$, the model hosts an exotic stripe phase~\cite{Chaloupka2010}. This phase spontaneously breaks spatial translational symmetry and is particularly intriguing within the Kitaev--Heisenberg model, as it emerges as an intermediate state between the conventional N\'eel antiferromagnetic phase and the exotic spin-liquid phase that preserves translational invariance. However, this stripe phase has so far remained elusive in quantum materials.

Allowing unequal lattice constants $a_x$, $a_y$, and $a_z$ gives rise to a rich phase diagram with distinct spin and translational symmetry breaking patterns. We perform exact diagonalization on a 24-site cluster with periodic boundary conditions to map out the phase diagram, as shown in Fig.~\ref{fig:2d}(b). The phase diagram is presented in the standard ternary form, which eliminates the overall energy scale: the center corresponds to the isotropic point, while the three vertices represent the limiting cases where only one of the $J_\gamma$'s is nonzero. 

A $\gamma$-stripe state simultaneously breaks two symmetries: the $\mathbb Z_2$ spin symmetry along the $\gamma$ direction, signaled by a nonzero $\langle S_\br^\gamma\rangle$, and the translational symmetry along the $\gamma$ axis, characterized by a nonzero $\langle S_\br^\gamma - S_{\br+\ba_\gamma}^\gamma\rangle$. The correlation between the direction of spin symmetry breaking and that of translational symmetry breaking directly manifests the spin--space locking at the quantum many-body level. At the isotropic point, all six stripe states (three choices of $\gamma$ times two symmetry-breaking branches) are degenerate. Introducing anisotropy lifts the degeneracy among the three $\gamma$ choices, and the smallest $J_\gamma$ selects the corresponding $\gamma$-stripe as the ground state. Conversely, when one coupling dominates, e.g., $J_x \gg J_y, J_z$, the system is governed by the $J_x$ bonds. In this limit, the many-body ground state is expected to be a direct product of $x$-bond singlets, restoring both spin and translational symmetries. The phase boundaries are determined by the maxima of the susceptibility of the stripe order parameter.

\textit{Conclusion.} To conclude, we have identified a \textit{magic} pair of principal quantum numbers where the electron-spin-dependent Rydberg interaction becomes significant. We emphasize two key features: (i) this interaction takes a simple universal form that is independent of the principal quantum numbers; and (ii) it exhibits spin--space-locked anisotropy. This distinctive property can be readily verified in current experiments by measuring quantum dynamics or energy spectroscopy with two atoms placed at different relative positions with respect to the spin quantization axis defined by an external magnetic field. This new interaction enriches the Rydberg quantum simulation toolbox. As a concrete application, we have proposed that two-layer arrays of such atoms can naturally realize the Kitaev--Heisenberg model, which hosts an unusual stripe phase that has not been previously observed.

\textit{Note Added.} The experimental signature of this prediction has been observed both from quantum dynamics~\cite{iop} and energy spectroscopy~\cite{apm} measurements recently.

\emph{Acknowledgments.} 
This work is supported by the Quantum Science and Technology-National Science and Technology Major Project under Grant No.~2025ZD0300400, and National Natural Science Foundation of China No. 12488301(H.Z.), No. U23A6004 (H.Z.), and No. 12504307 (C.L.). C.L. is also supported by the Tsinghua University Dushi program.


\begin{thebibliography}{99}
\bibitem{deltan1}
The case where $\Delta n = 1$ involves a principal quantum number exchange and will be discussed elsewhere.
\end{thebibliography}

\end{document}